Classical dense matter physics: some basic methods and results

Vladan Čelebonović

Institute of Physics, Pregrevica 118, 11080 Beograd- Zemun,Yugoslavia

vladan@phy.bg.ac.yu
vcelebonovic@sezampro.yu

Abstract: This is an introduction to the basic notions, some methods and open problems of dense matter physics and its applications in astrophysics. Experimental topics covered range from the work of P. W. Bridgman to the discovery and basic principles of use of the diamond anvil cell.On the theoretical side, the semiclassical method proposed by P.Savić and R.Kašanin is described. The choice of these methods is conditioned by their applicability in astrophysics and the author's research experience. The paper ends with a list of selected unsolved problems in dense matter physics and astrophysics, some (or all) of which could form the basis of future collaboration.

Introduction

The aim of this paper is to review some of the basic methods and results of classical dense matter physics. Classical here means loosely "not so dense that general-relativistic phenomena have to be taken into account". In practical terms, this corresponds to densities in the range $10^3$ - $10^5$ kg/m$^3$. The text is divided into several sections. The first one is devoted to the main instrument of modern static high pressure experiments-the diamond anvil cell (DAC). In the follow-up, a short review is given of the theoretical method developed by P.Savić and R.Kašanin in Belgrade, and the paper ends with a list of selected open problems in dense matter physics and astrophysics.

This paper has been written with the astrophysically oriented readers in mind. That is, people who are "at ease" in physics, but whose main interest are not the experimental or theoretical technicalities, but the applications to the interiors of various kinds of celestial objects.

Mentioning astrophysics as a science, one is inclined to think about vast regions of nearly empty interstellar or interplanetary space, scarcely populated by stars and other celestial objects. However, apart this aspect of astrophysics, there exists the opposite "end of story". In the interiors of planets, stars and various other kinds of astronomical objects, materials are subdued to extremely high values of the temperature and pressure. Although this has been known for decades, studies of dense matter physics have become experimentally possible only in the last 5-6 decades.

Static experiments

First attempts of studies of the behaviour of materials under high pressure have been made in the XVIII century [1] . It was noted at the time that a wealthy English "gentleman" named Mr.Canton compressed water at room temperature to a pressure of about 0.1 GPa. To his astonishement, water was transformed into ice. This was an isolated attempt, but first systematic studies of dense matter had to wait for another century.

They have been initiated by P.W.Bridgman at Harvard [2] ,who for his work won the Nobel prize in physics for 1946. Bridgman used large volume presses which had the advantage that they could contain large samples and had small P-T gradients. At the same time, they had the setback that the accessible region of P-T space was limited, and (what was perhaps worse) were expensive to build and maintain in operational conditions.

A real breakthrough in high pressure experimental work occurred around the middle of the XX century, with the invention of the diamond anvil cell (DAC). Details about this instrument are already available in the literature, and some have been discussed by the present author [3] and in the references given there.

A cross-section of a DAC is shown on the following figure,taken from [4].

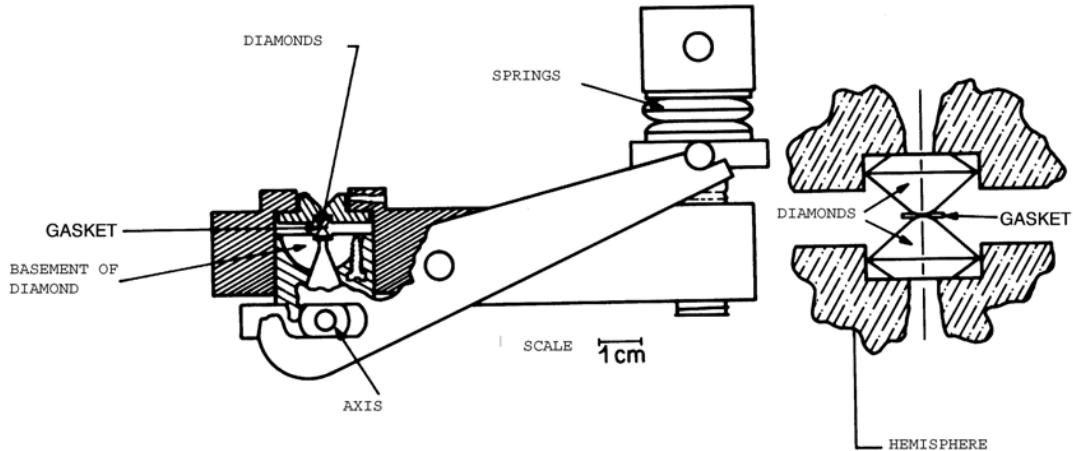

Figure 1. A cross section of the NBS type DAC

The term DAC comes from the fact that the most important part of this instrument is a couple of diamonds and a thin small metal plate (called the gasket) between them. Diamonds are important because they are hard and transparent; this implies that a specimen can be compressed to high pressure, and that it remains visible. This optical accessibility of specimen is a big advantage compared to Bridgman's experimental cells. A hole is drilled in the gasket, and in this way one gets a "working volume" in which the experiment is performed. Experiments in DACs are complicated by the miniaturized scale. Namely, diamonds are small. In order to use them as anvils, the gasket must be thin (around $250 \mu m$), and the diameter of the hole is approximately $200 \mu m$. Therefore specimen have to be small. Their typical size is about 40 $\mu m$ and they are usually prepared and inserted under a microscope. Apart the specimen, the working volume contains the pressure sensor and the pressure transmitting medium.

In the working volume, pressure is transmitted hydrostatically. The chemical composition of the hydrostatic fluid depends on the pressure and temperature at which the experiment is performed. For values of pressure smaller than 11 GPa, a mixture of methyl and ethyl alchocol in the ratio 4:1 is universally used [4].

The pressure is measured by the so-called "ruby scale". It has been shown several decades ago (for example [4] and references given there) that the spectrum of ruby ($Al_2O_3:Cr^{3+}$) excited with a laser beam or a spectral lamp consists of two lines with pressure dependent wavelengths. The physical mechanism in which these two lines originate is the transition $^4A_2 \to {^2E}$ in the ion $Cr^{3+}$ [5]. This pressure dependence is linear up to at least 30 GPa [4], while in its non-linear form the ruby scale can be applied up to 250 GPa; the final expression for the pressure dependence of the RI line (the stronger of the two) is [6]:

$$P[GPa] = 380.8\{(1+\Delta\lambda/694.2)^5 - 1\} \qquad (1)$$

The measurement of pressure is in fact indirect: the measured quantity is the change of wavelength of the RI line, and the pressure is calculated from eq.(1).

Experiments in DACs can be performed in the interval of temperatures between 4K and around 7000 K [3]. Record values of pressure measured so far are of the order of 450 GPa [7]. Note that this is of the order of magnitude of the pressure in the center of Earth.

The applicability of DACs in space science is virtually unlimited. It is well known that the interiors of planets and satellites are inaccessible to direct observation. Some of their observable parameters (such as, for example, the magnetic moment or the content of water as in [8]) critically depend on the conditions prevealing in their interiors. The only reliable method for investigating materials under such conditions is the use of DACs.

However, due to two "interfering complexities", that is the nature of phenomena occurring under high pressure and the actual methodology of experments under high pressure, drawing reliable conclusions about astrophysically relevant materials is far from easy. An excellent example, spanning half of century of research work and still unsolved, is offered by the behaviour of hydrogen under high pressure.

It has been predicted numerous times since the mid thirties of the last century that hydrogen becomes metallic at a pressure of the order of 250-300 GPa. These predictions were made many times, by different authors using differing methods, and all of them were giving nearly the same numerical value, which was taken more less as a measured value. For a long time values of pressure of this order were impossible to measure, but it was expected that once they become accessible to experiment this was going to be the result.

Two results of real experiments came as a complete surprise to the high pressure community. It was first shown by Ruoff *et.al.,* [9] that metallization of hydrogen does not occur for $P \le 342 GPa$. As a further set-back came the result from the Lawrence Livermore National Laboratory that the transition semiconductor$\to$ metallic in fluid hydrogen occurs at P=140 GPa, T=3000 K [10]. To the best of the author's knowledge the situation remains unsettled.

What can be concluded from this example? Hydrogen is a well known chemical element, extremely important in astrophysics. One would be inclined to think that everything is known about it, and that a phase transition point on its phase diagram can be predicted with near certainty. It appears that this is not the case, in spite of nearly 60 years of theoretical work on the problem. No definite explanation has been found, but in the opinion of the present author it should be sought in one of the following directions:

- Some of our theoretical ideas and methods are very probably wrong, or perhaps not applicable to hydrogen.
- Perhaps everything is right with our methodology, but we are making errors in predicting the experimentally measurable consequences of the metallization of hydrogen.

Settling of this "dispute" would have useful consequences for pure astronomy. Namely, hydrogen is a constituent of the giant planets of the solar system. Precise knowledge of its phase diagram under high pressure would enable getting more knowledge about their interiors. Extending this knowledge to high pressure and high temperatures could be useful for stellar astrophysics.

Experiments in DACs are capable of giving information about the behaviour of various materials in a large (but nevertheless limited) region of the P-T plane. A much wider region is accessible theoretically. The rest of this paper is devoted to a brief review of a semiclassical theory of dense matter proposed by P.Savić and R.Kašanin. As it has been discussed by the present author several times in recent years, ( for example [11] ) not many details will be given here.

Semiclassical studies of dense matter: a particular theory

A specimen of a solid, although it may appear as macroscopically small, is a typical example of a many-body system. It is a standard practice in statistical physics to describe the state of such a system by a Hamiltonian, which has the following general form:

$$H = \sum_{i=1}^{N}(-\frac{\hbar^2}{2m})\nabla_i^2 + \sum_{i=1}^{N}V(\vec{x}_i) + \sum_{i,j=1}^{N}v(|\vec{x}_i - \vec{x}_j|) \qquad (2)$$

The first term in this expression denotes the kinetic energy, the second is the interaction of the system with a possible external field, and the third denotes the pair-wise interaction of the particles. Now, obeying the rules of statistical mechanics, starting from the Hamiltonian one should calculate the free energy, and all the other thermodynamic potentials. Singularities in these potentials would be identified with the phase transition points in the system under consideration.

This algorithm may seem clear and straithforward, but actually it is far from being so. Namely, all the sums in eq.(2) go over all the particles in the system. As this number is usually of the order of Avogadro's number ($\sim 10^{23}$) clearly the summations in eq.(1) can no be performed in the arbitrary case.

However, this problem can be approached in a semi-classical manner, and that is what Pavle Savić and Radivoje Kašanin have managed to achieve.

At the beginning of the sixties, they have started developing a semiclassical theory of dense matter. Their starting point was astronomical: they have shown that the mean volumetric planetary densities can be related to the mean solar density by a simple relationship

$$\rho = \rho_0 2^\varphi \qquad (3)$$

where $\rho_0 = \dfrac{4}{3}$ is the mean solar density and $\varphi$ an integer. By choosing suitable values of this integer it becomes possible to reproduce the measured values of the densities. Using this result, and a fact known from geophysics and high pressure experiments that at certain values of pressure abrupt changes of the mass density occur, they came to the conclusion that atomic structure changes under the influence of high external pressure.

Starting from these ideas they proposed a set of 6 postulates which govern the behaviour of materials under high pressure. Each of them is based on known experimental results. Developing further these postulates, they have set up a calculational scheme, which gives the possibility of theoretical studies of dense matter physics, with applications in astronomy and laboratory studies. Full details about their postulates and algorithm are discussed in [11]. In the remainder of this section, we shall concentrate on a description of the applicability of this theory in astronomy.

Input data needed for modeling the internal structure of a planet, satellite or asteroid within this theory are the mass and radius of the object. Starting from these two values, it is possible to determine:

- the number of zones in the interior and their thickness
- the distribution of $P, \rho, T$ within the zones
- the magnetic moment of the object
- the mean atomic mass of the chemical mixture that the object is made of.
- the allowed interval in which the speed of axial rotation of the object must be.

Although it may seem overambitious, this list shows that within the theory proposed by P.Savić and R.Kašanin one can determine a complete model of the interior structure of a given planet, satellite or asteroid. Numerous examples exist in the literature (such as [11] and references given there) and the reader interested in details is advised to consult these publications.

All the planets except Saturn and Pluto, as well as the satellites of the Earth, Jupiter and Uranus and the asteroids 1 Ceres and 10 Hygiea have been modeled so far. Assembling the values of the mean atomic masses for all these objects, one gets the following table – which in fact shows the distribution of chemical elements in the present planetary system [11] .

Table I : the chemical composition of the Solar System

| object | <A> | satellite | <A> |
|---|---|---|---|
| Sun | 1.4 | Moon | 71 |
| Mercury | 113 | J1 | 70 |
| Venus | 28.12 | J2 | 71 |
| Earth | 26.56 | J3 | 18 |
| Mars | 69 | J4 | 19 |
| 1 Ceres | 96 | U1 | 38 |
| Jupiter | 1.55 | U2 | 43 |
| Saturn | / | U3 | 44 |
| Uranus | 6.5 | U4 | 32 |
| Neptune | 7.26 | U5 | 32 |
| Pluto | / | Triton | 67 |

Although this table is incomplete, several conclusions of physical interest can be drawn from it. It shows, for example, that the planetary system is inhomogenous chemically, and that the well known qualitative division between the terrestrial and jovian is reflected in their chemical composition. Striking similarities, which testify about violent events in the past history of the Solar System are visible from Table I. It turns out that Ceres and Mercury and Triton and Mars have similar values of the mean atomic mass. This can be interpreted as meaning that "once upon a time" these bodies originated in mutually close regions of the protosolar system, but that their orbits later diverged, perhaps due to collisions. Using the value of <A> calculated for 1 Ceres the mass of 10 Hygiea was calculated, and the result is in excellent agreement with the experimental value determined in celestial mechanics.

Values of gradients of <A> found in the jovian and uranian satellite systems have been interpreted as a consequence of various transport processes in the respective circum planetary accretion disks. The values of <A> found for the Earth and the Moon suggest that the Moon is a consequence of a "deep impact" into the Earth of a body which originated somewhere near the present orbit of Mars.

## What now?

In this paper we have described results in two branches of dense matter physics. The choice of the material was subjective – it was conditioned by the author's research experience, and by work performed in the author's laboratory. At the end, a few hints of possibilities of future work,and possibly collaborations are in order.

We have described the ruby scale for measuring high pressure. Altough it is universally used, it has its inherent problems. Research work is going on in laboratories around the world with the aim of finding new materials which could be pressure sensors. Work along these lines is also going on in the Institute of Physics ([12] an later work). Experiments along these lines could be performed jointly with colegues from Bulgaria.

At the time of preparation of this text, new results have emerged concerning hydrogen under high pressure [13] . They were obtained in US, in two big national laboratories (Sandia and Livermore), and concern the compressibility of hydrogen under high pressure. As it could be expected, the preliminary results released so far are contradictory. These results, although preliminary show that the problem of the behavior of hydrogen under high pressure is still open and unsolved. As such,it could be a possible subject for collaboration of physicists and astronomers in Bulgaria and Serbia. Taking into account the experimental facilities needed, it would have to be theoretical.

Another interesting line of research, actively pursued in Belgrade, concerns high pressure metrology. Basically, the problem is to find a replacement for the ruby scale. In view of the equipment which exists both in Bulgaria (ISSP) and Serbia ( Institute of Physics) a fruitful experimental collaboration could be initiated.

For pure astronomy, it could be interesting to measure optical reflection spectra and the behavior under high pressure of various materials which enter into the composition of the asteroids.

On the theoretical side,the theory proposed by P.Savić and R.Kašanin four decades ago needs a major "modernization" .For instance,work now in progress indicates that,according to modern data, eq.(3) should be replaced by

$$\rho = \frac{7}{5}\exp[\theta]$$

where the exponent is non-integral [14]. Details are currently being worked out,and the physical interpretation is (for the moment) unsolved.

The list of open problems in dense matter physics could be continued.The aim (and the ambition) is not to be exhaustive.Much simpler,the aim of this contribution is to present two basic notions, and indicate some open problems,which could form the basis for future joint research.